\documentclass[a4paper, 12pt]{article}

\usepackage{moreverb,url}

\usepackage[colorlinks,bookmarksopen,bookmarksnumbered,citecolor=red,urlcolor=red]{hyperref}
\usepackage{caption}
\usepackage{pdfpages}
\usepackage[tbtags]{amsmath}
\usepackage{filecontents}
\usepackage{authblk}

\PassOptionsToPackage{tbtags}{amsmath}

\newcommand\BibTeX{{\rmfamily B\kern-.05em \textsc{i\kern-.025em b}\kern-.08em
T\kern-.1667em\lower.7ex\hbox{E}\kern-.125emX}}

\DeclareMathOperator{\argmax}{arg\,max}

\usepackage[backend=bibtex,style=authoryear-comp,sorting=nyt,isbn=false,url=false, natbib=true, doi=false]{biblatex}
\begin{document}
\title{\textbf{Personalized Decision Making for Biopsies in Prostate Cancer Active Surveillance Programs}}

\author[1,*]{\small Anirudh Tomer}
\author[1]{Dimitris Rizopoulos}
\author[2]{Daan Nieboer}
\author[3]{Frank-Jan~Drost}
\author[3]{Monique J. Roobol}
\author[2,4]{Ewout W. Steyerberg}
\affil[1]{Department of Biostatistics, Erasmus University Medical Center, the Netherlands}
\affil[2]{Department of Public Health, Erasmus University Medical Center, the Netherlands}
\affil[3]{Department of Urology, Erasmus University Medical Center, the Netherlands}
\affil[4]{Department of Biomedical Data Sciences, Leiden University Medical Center, the Netherlands}
\affil[ ]{*\textit {email}: a.tomer@erasmusmc.nl}

\date{}
\maketitle

\begin{abstract}
\textbf{Background.} Low-risk prostate cancer patients enrolled in active surveillance programs commonly undergo biopsies for examination of cancer progression. Biopsies are conducted as per a fixed and frequent schedule (e.g., annual biopsies). Since biopsies are burdensome, patients do not always comply with the schedule, which increases the risk of delayed detection of cancer progression.

\textbf{Objective.} Our aim is to better balance the number of biopsies (burden) and the delay in detection of cancer progression (less is beneficial), by personalizing the decision of conducting biopsies.

\textbf{Data Sources.} We use patient data of the world's largest active surveillance program (PRIAS). It enrolled 5270 patients, had 866 cancer progressions, and an average of nine prostate-specific antigen (PSA) and five digital rectal examination (DRE) measurements per patient.

\textbf{Methods.} Using joint models for time-to-event and longitudinal data, we model the historical DRE and PSA measurements, and biopsy results of a patient at each follow-up visit. This results in a visit and patient-specific cumulative risk of cancer progression. If this risk is above a certain threshold, we schedule a biopsy. We compare this personalized approach with the currently practiced biopsy schedules via an extensive and realistic simulation study, based on a replica of the patients from the PRIAS program.

\textbf{Results.} The personalized approach saved a median of six biopsies (median:~4,~IQR:~2--5), compared to the annual schedule (median:~10,~IQR:~3--10). However, the delay in detection of progression (years) is similar for the personalized (median:~0.7,~IQR:~0.3--1.0) and the annual schedule (median:~0.5,~IQR:~0.3--0.8).

\textbf{Conclusions.} We conclude that personalized schedules provide substantially better balance in the number of biopsies per detected progression for men with low-risk prostate cancer.
\end{abstract}

\section{Introduction}
\label{sec:introduction}
Prostate cancer is the second most frequently diagnosed cancer in men worldwide \citep{GlobalCancerStats2012}. In prostate cancer screening programs, many of the diagnosed tumors are clinically insignificant (over-diagnosed) \citep{etzioni2002overdiagnosis}. To avoid further over-treatment, patients diagnosed with low-grade prostate cancer are commonly advised to join active surveillance (AS) programs. In AS, invasive treatments such as surgery are delayed until cancer progresses. Cancer progression is routinely monitored via serum prostate-specific antigen (PSA) measurements, a protein biomarker; digital rectal examination (DRE) measurements, a measure of the size and location of the tumor; and biopsies.

While larger values for PSA and/or DRE, may indicate cancer progression, biopsies are the most reliable cancer progression examination technique used in AS. When a patient's biopsy Gleason score becomes larger than 6 (positive biopsy, cancer progression detected), AS is stopped and the patient is advised treatment \citep{bokhorst2015compliance}. However, biopsies are invasive, painful, and prone to medical complications \citep{ehdaie2014impact,fujita2009serial}. Hence, they are conducted intermittently until a positive biopsy. Consequently, at the time of a positive biopsy, cancer progression may be observed with a delay of unknown duration. This delay is defined as the difference between the time of the positive biopsy and the unobserved true time of cancer progression. Thus, the decision to conduct biopsies requires a compromise between the burden of biopsy and the potential delay in the detection of cancer progression.

In AS, a delay in the detection of cancer progression around 12 to 14 months is assumed to be unlikely to substantially increase the risk of adverse downstream outcomes \citep{inoue2018comparative,carvalho}. However, for biopsies, there is little consensus on the time gap between them \citep{loeb2014heterogeneity,bruinsma2016active,nieboer2018active}. Many AS programs focus on minimizing the delay in the detection of cancer progression, by scheduling biopsies annually for all patients. A drawback of annual biopsies, and other currently practiced fixed/heuristic schedules \citep{loeb2014heterogeneity,bruinsma2016active,nieboer2018active}, is that they ignore the large variation in the time of cancer progression of AS patients. While they may work well for patients who progress early (\textit{fast progressing}) in AS, but for a large proportion of patients who do not progress, or progress late (\textit{slow progressing}) in AS, many unnecessary burdensome biopsies are scheduled. To mediate the burden between the \textit{fast} and \textit{slow progressing} patients, the world's largest AS program, Prostate Cancer Research International Active Surveillance \citep{bokhorst2016decade} (PRIAS), schedules annual biopsies only for patients with a low PSA doubling time \citep{bokhorst2015compliance}. For everyone else, PRIAS schedules biopsies at following fixed follow-up times: year one, four, seven, and ten, and every five years thereafter. Despite this effort in PRIAS, patients may get scheduled for four to ten biopsies over a period of ten years. Therefore, compliance for biopsies is low in PRIAS \citep{bokhorst2015compliance}. This can lead to a delay in the detection of cancer progression, and reduce the effectiveness of AS.

\begin{figure}[!htb]
\captionsetup{justification=justified}
\centerline{\includegraphics[width=\columnwidth]{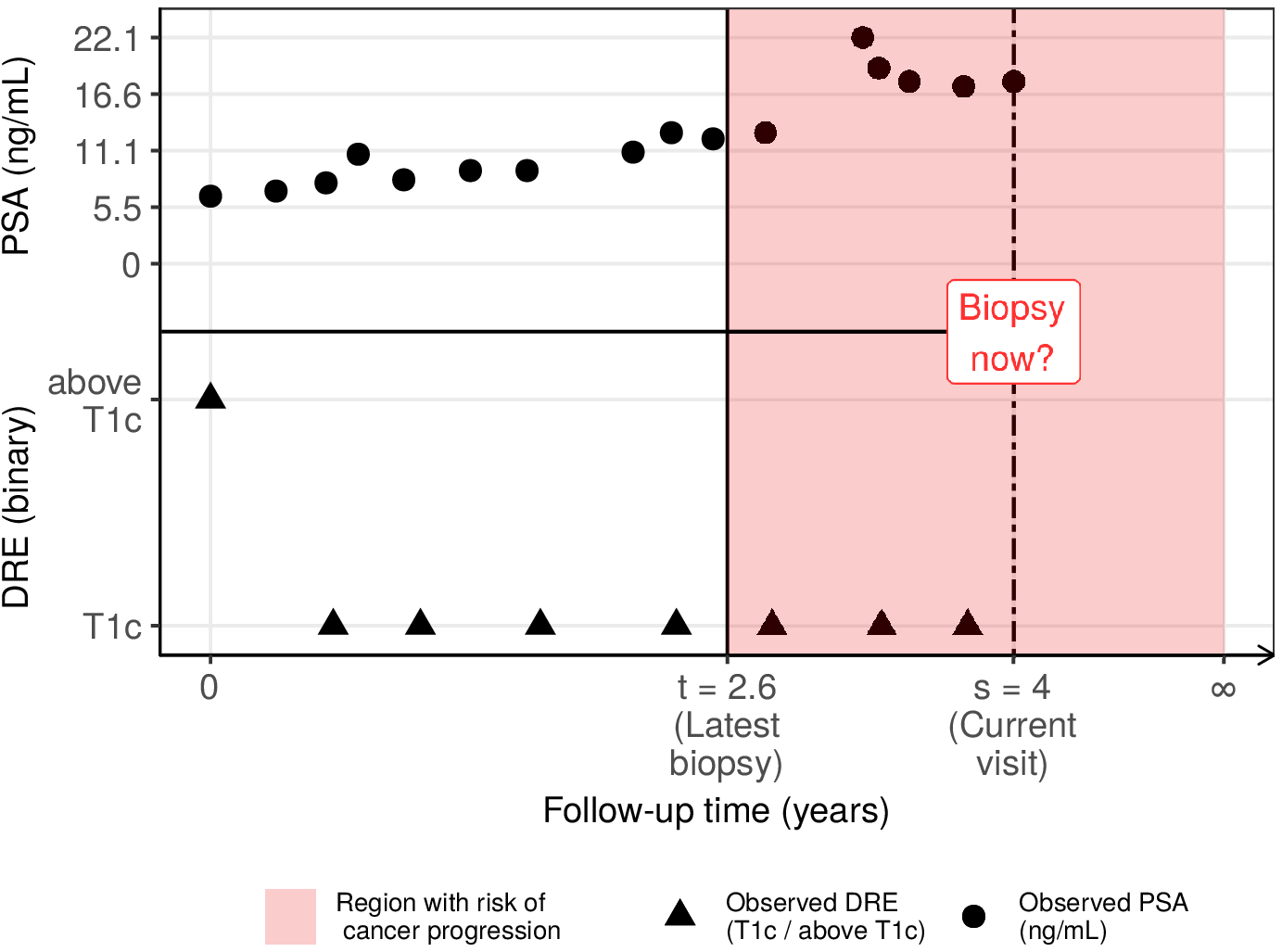}}
\input{Figure1Caption}
\label{Figure1}
\end{figure}

We aim to better balance the number of biopsies (more are burdensome), and the delay in the detection of cancer progression (less is beneficial), than currently practiced schedules. We intend to achieve this by personalizing the decision to conduct biopsies (see Figure~\ref{Figure1}). These decisions are made at a patient's pre-scheduled follow-up visits for DRE and PSA measurements. To develop the personalized decision making methodology, we utilize the data of the patients enrolled in the PRIAS study. We model this data and develop the personalized approach using joint models for time-to-event and longitudinal data \citep{tsiatis2004joint,rizopoulos2012joint}. In order to compare the personalized approach with current schedules, we conduct an extensive simulation study based on a replica of the patients from the PRIAS program. 

\section{Methods}
\label{sec:methods}
\subsection{Study Population}
\label{subsec:study_population}
\begin{figure}[!htb]
\captionsetup{justification=justified}
\centerline{\includegraphics[width=\columnwidth]{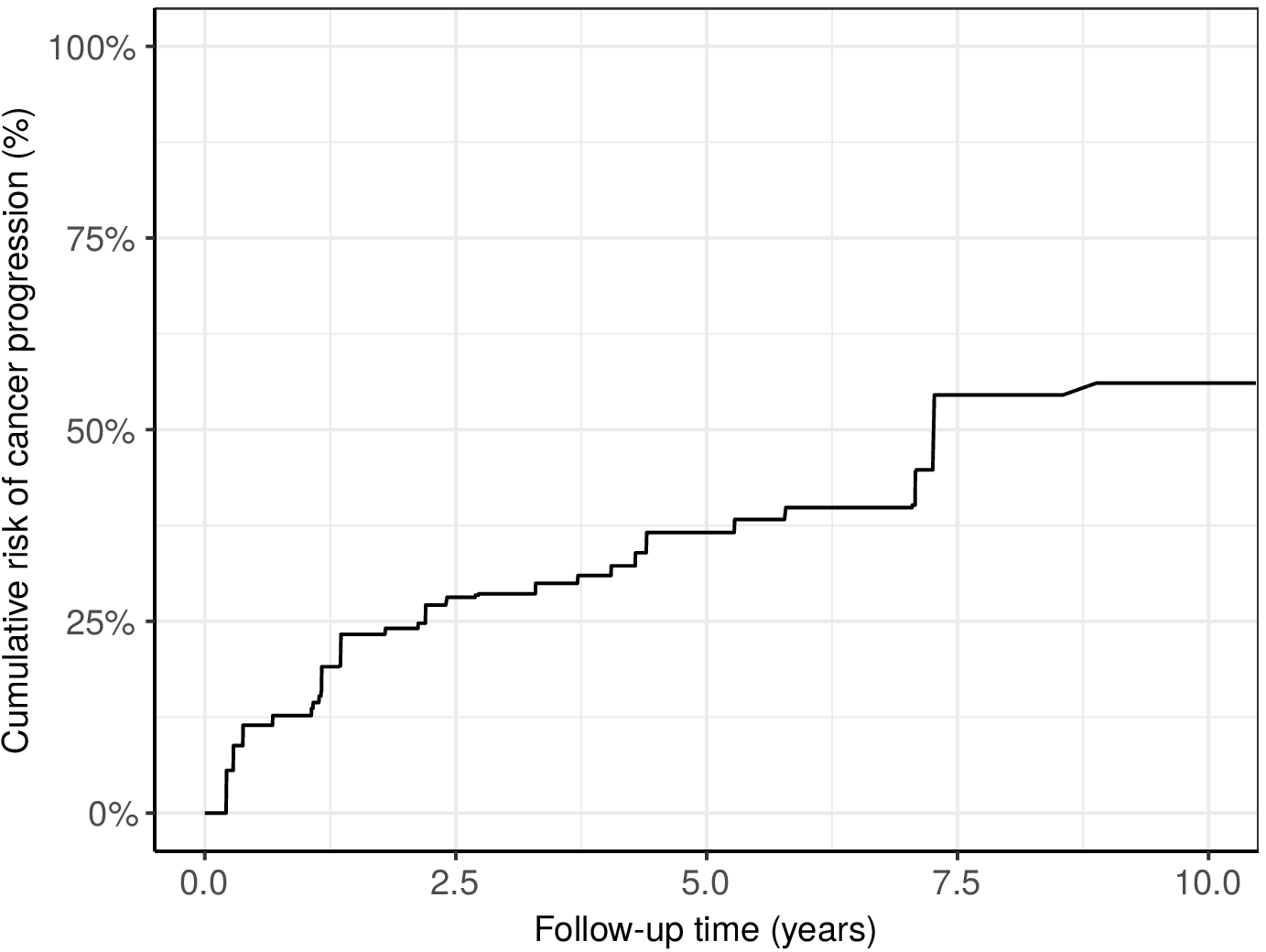}}
\input{Figure2Caption}
\label{Figure2}
\end{figure}

To develop our methodology we use the data of prostate cancer patients from the world's largest AS study called PRIAS \citep{bokhorst2016decade} (see Table~\ref{Table1}). More than 100 medical centers from 17 countries worldwide contribute to the collection of data, utilizing a common study protocol and a web-based tool, both available at \url{www.prias-project.org}. We use data collected over a period of ten years, between December 2006 (beginning of PRIAS study) and December 2016. The primary event of interest is cancer progression detected upon a positive biopsy. The time of cancer progression is interval censored because biopsies are scheduled periodically. Biopsies are scheduled as per the PRIAS protocol (see \hyperref[sec:introduction]{Introduction}). There are three types of competing events, namely death, removal of patients from AS on the basis of their observed DRE and PSA measurements, and loss to follow-up. We assume these three types of events to be censored observations (see Appendix~A.5 for details). However, our model allows removal of patients to depend on observed longitudinal data and baseline covariates of the patient. Under the aforementioned assumption of censoring, Figure~\ref{Figure2} shows the cumulative risk of cancer progression over the study follow-up period.

\begin{table}
\captionsetup{justification=justified}
\small\sf\centering
\caption{\textbf{Summary statistics for the PRIAS dataset}. The primary event of interest is cancer progression. A DRE measurement equal to T1c\cite{schroder1992tnm} indicates a clinically inapparent tumor which is not palpable or visible by imaging, while tumors with $\mbox{DRE} > \mbox{T1c}$ are palpable. The abbreviation IQR means interquartile range.}
\label{Table1}
\begin{tabular}{lr}
\hline
\hline
Data & Value\\
\hline
Total patients & 5270\\
Cancer progression (primary event) & 866\\
Loss to follow-up (anxiety or unknown) & 685\\
Removal on the basis of PSA and DRE & 464\\
Death (unrelated to prostate cancer) & 61\\
Death (related to prostate cancer) & 2\\
\hline
Median Age (years) & 70 (IQR: 65--75)\\
Total PSA measurements & 46015\\
Median number of PSA per patient &  7 (IQR: 7--12)\\
Median PSA value (ng/mL) & 5.6 (IQR: 4.0--7.5)\\
Total DRE measurements & 25606\\
Median number of DRE per patient & 4 (IQR: 3--7)\\
$\mbox{DRE} = \mbox{T1c}$ (\%) & 23538/25606 (92\%) \\
\hline
\end{tabular}
\end{table}

For all patients, PSA measurements (ng/mL) are scheduled every 3 months for the first 2 years and every 6 months thereafter. The DRE measurements are scheduled every 6 months. We use the DRE measurements as ${\mbox{DRE} = \mbox{T1c}}$ versus $\mbox{DRE} > \mbox{T1c}$. A DRE measurement equal to T1c \citep{schroder1992tnm} indicates a clinically inapparent tumor which is not palpable or visible by imaging, while tumors with $\mbox{DRE} > \mbox{T1c}$ are palpable.

\textbf{Data Accessibility:} The PRIAS database is not openly accessible. However, access to the database can be requested on the basis of a study proposal approved by the PRIAS steering committee. The website of the PRIAS program is \url{www.prias-project.org}.

\subsection{A Bivariate Joint Model for the Longitudinal PSA, and DRE Measurements, and Time of Cancer Progression}
Let $T_i^*$ denote the true cancer progression time of the ${i\mbox{-th}}$ patient included in PRIAS. Since biopsies are conducted periodically, $T_i^*$ is observed with interval censoring ${l_i < T_i^* \leq r_i}$. When progression is observed for the patient at his latest biopsy time $r_i$, then $l_i$ denotes the time of the second latest biopsy. Otherwise, $l_i$ denotes the time of the latest biopsy and ${r_i=\infty}$. Let $\boldsymbol{y}_{di}$ and $\boldsymbol{y}_{pi}$ denote his observed DRE and PSA longitudinal measurements, respectively. The observed data of all $n$ patients is denoted by ${\mathcal{D}_n = \{l_i, r_i, \boldsymbol{y}_{di}, \boldsymbol{y}_{pi}; i = 1, \ldots, n\}}$.

\begin{figure}[!htb]
\captionsetup{justification=justified}
\centerline{\includegraphics[width=\columnwidth]{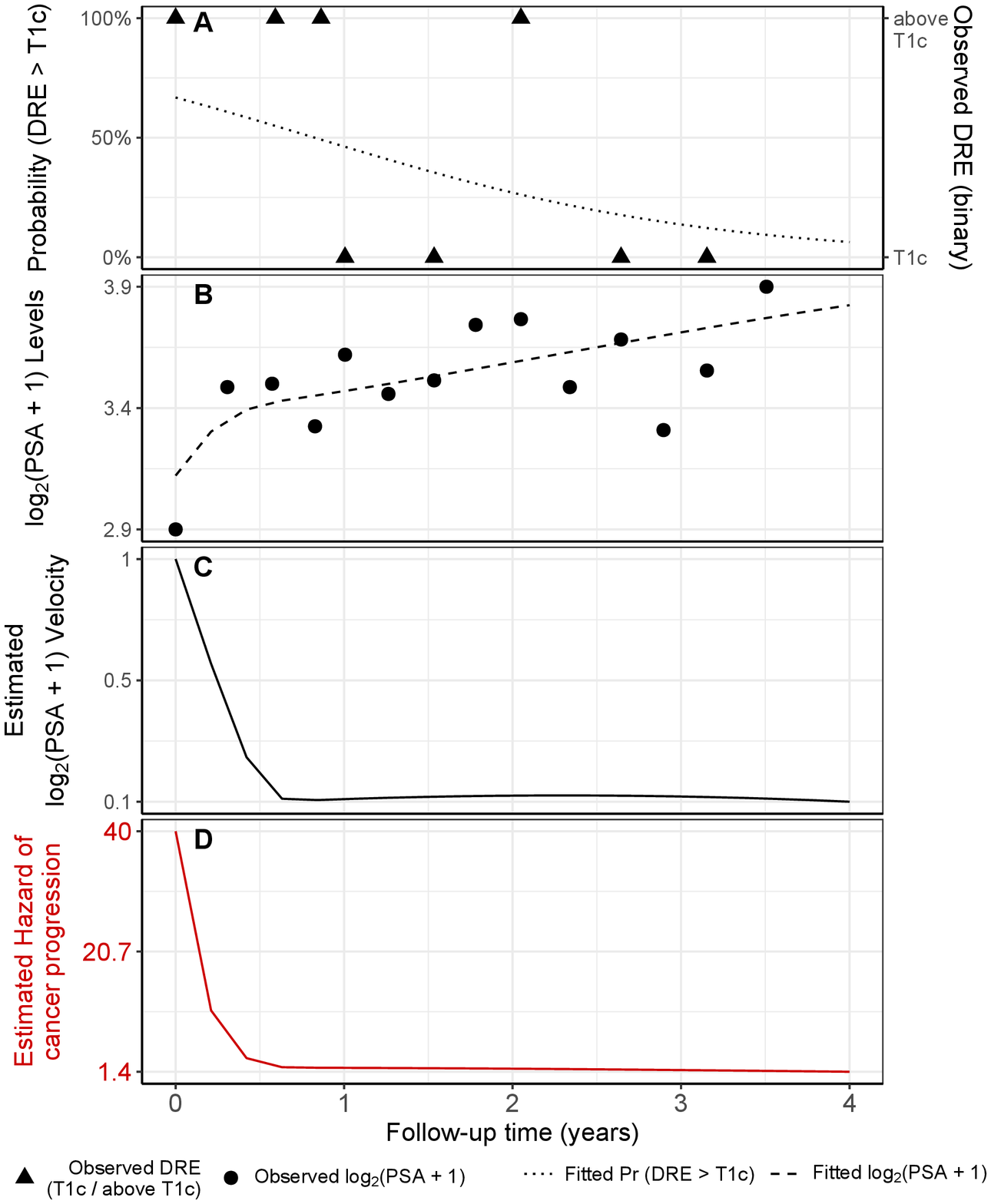}}
\input{Figure3Caption}
\label{Figure3}
\end{figure}

In our joint model, the patient-specific DRE and PSA measurements over time are modeled using a bivariate generalized linear mixed effects sub-model. The sub-model for DRE is given by (see~Panel~A, Figure~\ref{Figure3}):
\begin{equation}
\label{eq:long_model_dre}
\begin{split}
    \mbox{logit} \big[\mbox{Pr}\{y_{di}(t) > \mbox{T1c}\}\big] &= \beta_{0d} + b_{0di} + (\beta_{1d} + b_{1di}) t\\
    &+ \beta_{2d} (\mbox{Age}_i-70)\\ & + \beta_{3d} (\mbox{Age}_i-70)^2
    \end{split}
\end{equation}
where, $t$ denotes the follow-up visit time, and $\mbox{Age}_i$ is the age of the ${i\mbox{-th}}$ patient at the time of inclusion in AS. We have centered the Age variable around the median age of 70 years for better convergence during parameter estimation. However, this does not change the interpretation of the parameters corresponding to the Age variable. The fixed effect parameters are denoted by ${\{\beta_{0d}, \ldots, \beta_{3d}\}}$, and ${\{b_{0di}, b_{1di}\}}$ are the patient specific random effects. With this definition, we assume that the patient-specific log odds of obtaining a DRE measurement larger than T1c remain linear over time. 

The mixed effects sub-model for PSA is given by (see~Panel~B, Figure~\ref{Figure3}):
\begin{equation}
\label{eq:long_model_psa}
\begin{split}
    \log_2 \big\{y_{pi}(t) + 1\big\} &= m_{pi}(t) + \varepsilon_{pi}(t),\\
    m_{pi}(t) &= \beta_{0p} + b_{0pi} + \sum_{k=1}^4 (\beta_{kp} + b_{kpi})  B_k(t,\mathcal{K})\\ 
    &+ \beta_{5p} (\mbox{Age}_i-70) + \beta_{6p} (\mbox{Age}_i-70)^2,
    \end{split}
\end{equation}
where, $m_{pi}(t)$ denotes the underlying measurement error free value of $\log_2 (\mbox{PSA} + 1)$ transformed \citep{pearson1994mixed,lin2000latent} measurements at time $t$. We model it non-linearly over time using B-splines \citep{de1978practical}. To this end, our B-spline basis function $B_k(t, \mathcal{K})$ has 3 internal knots at $\mathcal{K} = \{0.1, 0.7, 4\}$ years, and boundary knots at 0 and 5.42 years (95-th percentile of the observed follow-up times). This specification allows fitting the $\log_2 (\mbox{PSA} + 1)$ levels in a piecewise manner for each patient separately. The internal and boundary knots specify the different time periods (analogously pieces) of this piecewise nonlinear curve. The fixed effect parameters are denoted by ${\{\beta_{0p},\ldots,\beta_{6p}\}}$, and ${\{b_{0pi}, \ldots, b_{4pi}\}}$ are the patient specific random effects. The error $\varepsilon_{pi}(t)$ is assumed to be t-distributed with three degrees of freedom (see~Appendix~B.1) and scale $\sigma$, and is independent of the random effects. 

To account for the correlation between the DRE and PSA measurements of a patient, we link their corresponding random effects. More specifically, the complete vector of random effects ${\boldsymbol{b}_i = (b_{0di}, b_{1di}, b_{0pi}, \ldots, b_{4pi})^T}$ is assumed to follow a multivariate normal distribution with mean zero and variance-covariance matrix $\boldsymbol{D}$.

To model the impact of DRE and PSA measurements on the risk of cancer progression, our joint model uses a relative risk sub-model. More specifically, the hazard of cancer progression $h_i(t)$ at a time $t$ is given by (see~Panel~D, Figure~\ref{Figure3}):
\begin{equation}
\label{eq:rel_risk_model}
\begin{split}
    h_i(t) &= h_0(t) \exp\Big(\gamma_1 (\mbox{Age}_i-70) + \gamma_2 (\mbox{Age}_i-70)^2\\
    &+\alpha_{1d} \mbox{logit} \big[\mbox{Pr}\{y_{di}(t) > \mbox{T1c}\}\big]\\&+ \alpha_{1p} m_{pi}(t) + \alpha_{2p} \frac{\partial m_{pi}(t)}{\partial {t}}\Big),
    \end{split}
\end{equation}
where, $\gamma_1, \gamma_2$ are the parameters for the effect of age. The parameter $\alpha_{1d}$ models the impact of log odds of obtaining a $\mbox{DRE} > \mbox{T1c}$ on the hazard of cancer progression. The impact of PSA on the hazard of cancer progression is modeled in two ways: a) the impact of the error free underlying PSA value $m_{pi}(t)$ (see~Panel~B, Figure~\ref{Figure3}), and b) the impact of the underlying PSA velocity $\partial m_{pi}(t)/\partial {t}$ (see~Panel~C, Figure~\ref{Figure3}). The corresponding parameters are $\alpha_{1p}$ and $\alpha_{2p}$, respectively. Lastly, $h_0(t)$ is the baseline hazard at time $t$, and is modeled flexibly using P-splines \citep{eilers1996flexible}. The detailed specification of the baseline hazard $h_0(t)$, and the joint parameter estimation of the two sub-models using the Bayesian approach (R package \textbf{JMbayes}, see \cite{rizopoulosJMbayes}) are presented in Appendix A of the supplementary material.

\subsection{Personalized Decisions for Biopsy}
\label{subsec:pers_decision_making}
\begin{figure}[!htb]
\captionsetup{justification=justified}
\centerline{\includegraphics[width=\columnwidth]{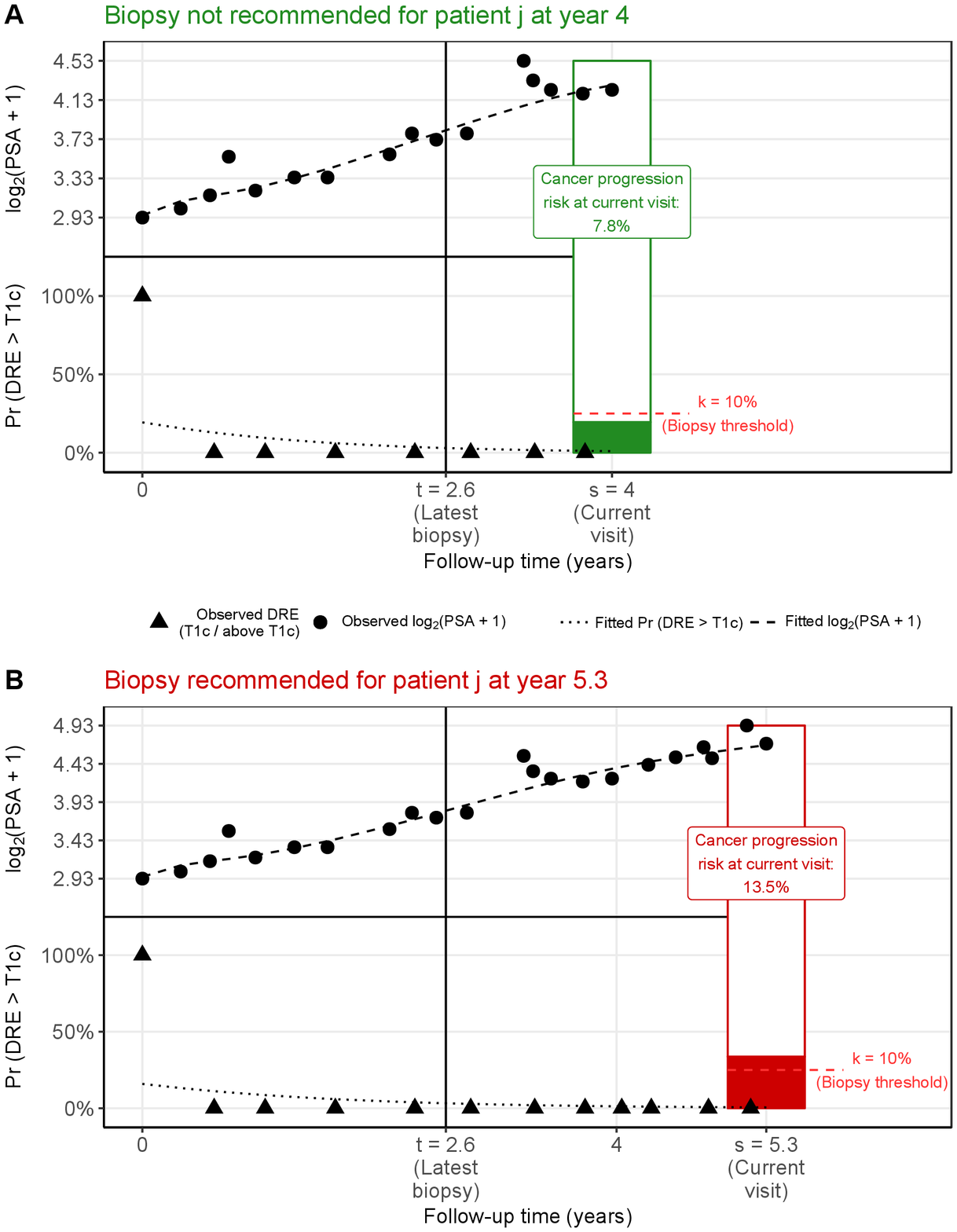}}
\input{Figure4Caption}
\label{Figure4}
\end{figure}

Let us assume that a decision of conducting a biopsy is to be made for a new patient $j$ shown in Figure~\ref{Figure1}, at his current follow-up visit time $s$. Let $t\leq s$ be the time of his latest negative biopsy. Let $\mathcal{Y}_{dj}(s)$ and $\mathcal{Y}_{pj}(s)$ denote his observed DRE and PSA measurements up to the current visit, respectively. From the observed measurements we want to extract the underlying measurement error free trend of $\log_2 (\mbox{PSA} + 1)$ values and velocity, and the log odds of obtaining $\mbox{DRE} > \mbox{T1c}$. We intend to combine them to inform us when the cancer progression is to be expected (see~Figure~\ref{Figure4}), and to further guide the decision making on whether to conduct a biopsy at the current follow-up visit. The combined information is given by the following posterior predictive distribution $g(T^*_j)$ of his time of cancer progression $T^*_j > t$ (see~Appendix~A.4 for details):
\begin{equation}
\label{eq:post_pred_dist}
g(T^*_j) = p\big\{T^*_j \mid T^*_j > t, \mathcal{Y}_{dj}(s), \mathcal{Y}_{pj}(s), \mathcal{D}_n\big\}.
\end{equation}
The distribution $g(T^*_j)$ is not only patient-specific, but also updates as extra information is recorded at future follow-up visits.

A key ingredient in the decision of conducting a biopsy for patient $j$ at the current follow-up visit time $s$ is the personalized cumulative risk of observing a cancer progression at time $s$ (illustrated in Figure~\ref{Figure4}). This risk can be derived from the posterior predictive distribution $g(T^*_j)$ \citep{rizopoulos2011dynamic}, and for $s \geq t$ it is given by:
\begin{equation}
\label{eq:dynamic_risk_prob}
R_j(s \mid t) = \mbox{Pr}\big\{T^*_j \leq s \mid T^*_j > t, \mathcal{Y}_{dj}(s), \mathcal{Y}_{pj}(s), \mathcal{D}_n\big\}.
\end{equation}
A simple and straightforward approach to decide upon conducting a biopsy for patient $j$ at the current follow-up visit would be to do so if his personalized cumulative risk of cancer progression at the visit is higher than a certain threshold $0 \leq \kappa \leq 1$. For example, as shown in Panel~B of Figure~\ref{Figure4}, biopsy at a visit may be scheduled if the personalized cumulative risk is higher than 10\% (example risk threshold). This decision making process is iterated over the follow-up period, incorporating on each subsequent visit the newly observed data, until a positive biopsy is observed. Subsequently, an entire personalized schedule of biopsies for each patient can be obtained.

The choice of the risk threshold dictates the schedule of biopsies and has to be made on each subsequent follow-up visit of a patient. In this regard, a straightforward approach is choosing a fixed risk threshold, such as 5\% or 10\% risk, at all follow-up visits. Fixed risk thresholds may be chosen by patients and/or doctors according to how they weigh the relative harms of doing an unnecessary biopsy versus a missed cancer progression (e.g., 10\% threshold means a 1:9 ratio) if the biopsy is not conducted \citep{vickers2006decision}. An alternative approach is that at each follow-up visit a unique threshold is chosen on the basis of its classification accuracy. More specifically, given the time of latest biopsy $t$ of patient $j$, and his current visit time $s$\, we find a visit-specific biopsy threshold $\kappa$, which gives the highest cancer progression detection rate (true positive rate, or TPR) for the period $(t, s]$. However, we also intend to balance for unnecessary biopsies (high false positive rate), or a low number of correct detections (high false negative rate) when the false positive rate is minimized. An approach to mitigating these issues is to maximize the TPR and positive predictive value (PPV) simultaneously. To this end, we utilize the $\mbox{F}_1$ score, which is a composite of both TPR and PPV (estimated as in Rizopoulos~et~al., 2017 \cite{landmarking2017}), and is defined as: 
\begin{equation}
\label{eq:F1_TPR_PPV}
\begin{split}
\mbox{F}_1(t,  s, \kappa) &= 2\frac{\mbox{TPR}(t,  s, \kappa)\ \mbox{PPV}(t,  s, \kappa)}{\mbox{TPR}(t,  s, \kappa) + \mbox{PPV}(t,  s, \kappa)},\\
\mbox{TPR}(t,  s, \kappa) &= \mbox{Pr}\big\{R_j(s \mid t) > \kappa \mid t < T^*_j \leq s\big\},\\
\mbox{PPV}(t,  s, \kappa) &= \mbox{Pr}\big\{t < T^*_j \leq s \mid R_j(s \mid t) > \kappa \big\},
\end{split}
\end{equation}
where, $\mbox{TPR}(t,  s, \kappa)$ and $\mbox{PPV}(t,  s, \kappa)$ are the time dependent true positive rate and positive predictive value, respectively. These values are unique for each combination of the time period $(t, s]$ and the risk threshold $\kappa$ that is used to discriminate between the patients whose cancer progresses in this time period versus the patients whose cancer does not progress. The same holds true for the resulting $\mbox{F}_1$ score denoted by $\mbox{F}_1(t,  s, \kappa)$. The $\mbox{F}_1$ score ranges between 0 and 1, where a value equal to 1 indicates perfect TPR and PPV. Thus the highest $\mbox{F}_1$ score is desired in each time period $(t, s]$. This can be achieved by choosing a risk threshold $\kappa$ which maximizes $\mbox{F}_1(t, s, \kappa)$. That is, during a patient's visit at time $s$, given that his latest biopsy was at time $t$, the visit-specific risk threshold to decide a biopsy is given by ${\kappa=\argmax_{\kappa} \mbox{F}_1(t, s, \kappa)}$. The criteria on which we evaluate the personalized schedules based on fixed and visit-specific risk thresholds is the total number of biopsies scheduled, and the delay in detection of cancer progression (details in \hyperref[sec:results]{Results}). 

\subsection{Simulation Study}
\label{subsec:sim_study}
Although the personalized decision making approach is motivated by the PRIAS study, it is not possible to evaluate it directly on the PRIAS dataset. This is because the patients in PRIAS have already had their biopsies as per the PRIAS protocol. In addition, the true time of cancer progression is interval or right censored for all patients, making it impossible to correctly estimate the delay in detection of cancer progression due to a particular schedule. To this end, we conduct an extensive simulation study to find the utility of personalized, PRIAS, and fixed/heuristic schedules. For a realistic comparison, we simulate patient data from the joint model fitted to the PRIAS dataset. The simulated population has the same ten year follow-up period as the PRIAS study. In addition, the estimated relations between DRE and PSA measurements, and the risk of cancer progression are retained in the simulated population.

From this population, we first sample 500 datasets, each representing a hypothetical AS program with 1000 patients in it. We generate a true cancer progression time for each of the ${\mbox{500} \times \mbox{1000}}$ patients and then sample a set of DRE and PSA measurements at the same follow-up visit times as given in PRIAS protocol. We then split each dataset into training (750 patients) and test (250 patients) parts, and generate a random and non‐informative censoring time for the training patients. We next fit a joint model of the specification given in Equations (\ref{eq:long_model_dre}), (\ref{eq:long_model_psa}), and (\ref{eq:rel_risk_model}) to each of the 500 training datasets and obtain MCMC samples from the 500 sets of the posterior distribution of the parameters. 

In each of the 500 hypothetical AS programs, we utilize the corresponding fitted joint models to develop cancer progression risk profiles for each of the ${\mbox{500} \times \mbox{250}}$ test patients. We make the decision of biopsies for patients at their pre-scheduled follow-up visits for DRE and PSA measurements (see \hyperref[subsec:study_population]{Study Population}), on the basis of their estimated personalized cumulative risk of cancer progression. These decisions are made iteratively until a positive biopsy is observed. A recommended gap of one year between consecutive biopsies \citep{bokhorst2015compliance} is also maintained. Subsequently, for each patient, an entire personalized schedule of biopsies is obtained.

We evaluate and compare both personalized and currently practiced schedules of biopsies in this simulation study. Comparison of the schedules is based on the number of biopsies scheduled and the corresponding delay in the detection of cancer progression. We evaluate the following currently practiced fixed/heuristic schedules: biopsy annually, biopsy every one and a half years, biopsy every two years and biopsy every three years. We also evaluate the biopsy schedule of the PRIAS program (see \hyperref[sec:introduction]{Introduction}). For the personalized biopsy schedules, we evaluate schedules based on three fixed risk thresholds: 5\%, 10\%, and 15\%, corresponding to a missed cancer progression being 19, 9, and 5.5 times more harmful than an unnecessary biopsy \citep{vickers2006decision}, respectively. We also implement a personalized schedule where for each patient, visit-specific risk thresholds are chosen using $\mbox{F}_1$ score.

\section{Results}
\label{sec:results}
From the joint model fitted to the PRIAS dataset, we found that both $\log_2 \{\mbox{PSA} + 1\}$ velocity,  and log odds of having $\mbox{DRE} > \mbox{T1c}$  were significantly associated with the hazard of cancer progression. For any patient, an increase in $\log_2 \{\mbox{PSA} + 1\}$ velocity from -0.03 to 0.16 (first and third quartiles of the fitted velocities, respectively) corresponds to a 1.94 fold increase in the hazard of cancer progression. Whereas, an increase in odds of $\mbox{DRE} > \mbox{T1c}$ from -6.650 to -4.356 (first and third quartiles of the fitted log odds, respectively) corresponds to a 1.40 fold increase in the hazard of cancer progression. Detailed results pertaining to the fitted joint model are presented in Appendix B.

\subsection{Comparison of Various Approaches for Biopsies}
\begin{figure}[!htb]
\captionsetup{justification=justified}
\centerline{\includegraphics[width=\columnwidth]{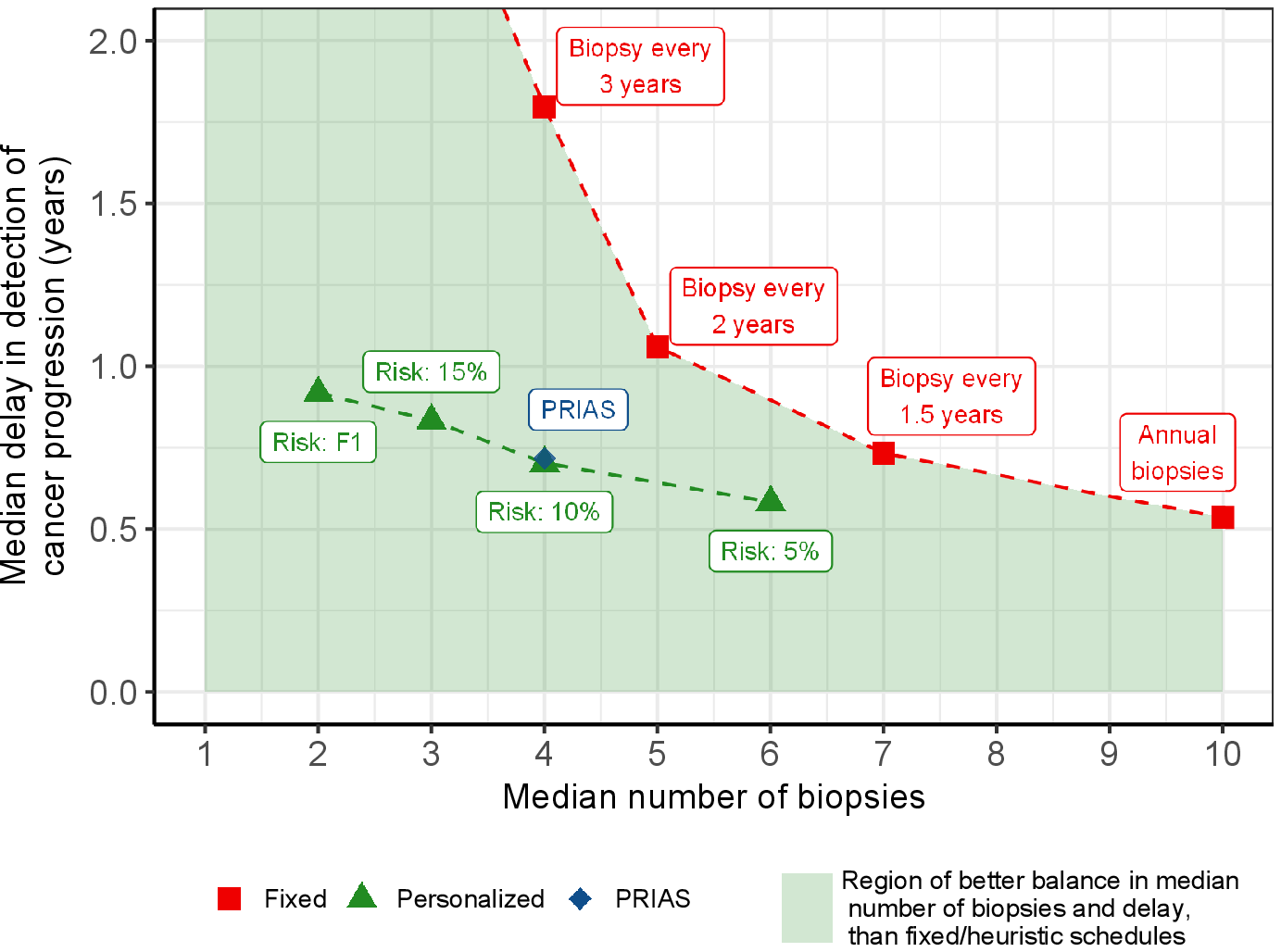}}
\input{Figure5Caption}
\label{Figure5}
\end{figure}

\begin{figure}[!htb]
\captionsetup{justification=justified}
\centerline{\includegraphics[width=\columnwidth]{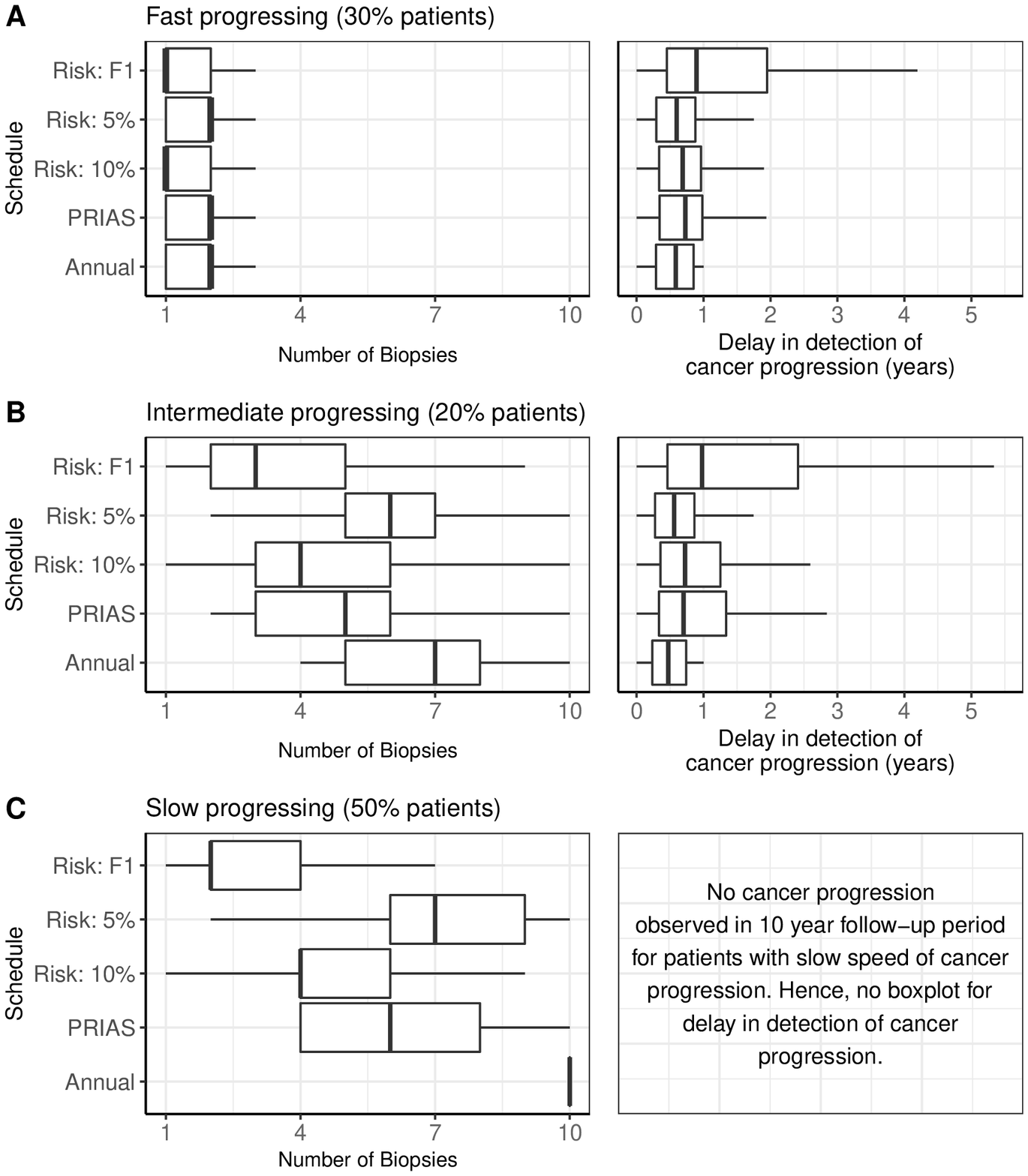}}
\input{Figure6Caption}
\label{Figure6}
\end{figure}

From the simulation study, we obtain the number of biopsies and the delay in detection of cancer progression for each of the ${\mbox{500} \times \mbox{250}}$ test patients using different schedules. Figure~\ref{Figure5} shows that the personalized and PRIAS approaches fall in the region of better balance between the median number of biopsies and the median delay than fixed/heuristic schedules. We next evaluate these schedules on the basis of both median and interquartile range (IQR) of the number of biopsies and delay (see Figure~\ref{Figure6}). For brevity, only the most widely used annual and PRIAS schedules, the proposed personalized approach with fixed risk thresholds of 5\% and 10\%, and visit-specific threshold chosen using $\mbox{F}_1$ score are discussed next (see~Appendix~C for remaining).

Since patients have varying cancer progression speeds, the impact of each schedule also varies with it. In order to highlight these differences, we divide results for three types of patients, as per their time of cancer progression. They are \textit{fast, intermediate,} and \textit{slow progressing} patients. Although such a division may be imperfect and can only be done retrospectively in a simulation setting, we show results for these three groups for the purpose of illustration. Roughly 50\% of the patients did not obtain cancer progression in the ten year follow-up period of the simulation study. We assume these patients to be \textit{slow progressing} patients. We assume \textit{fast progressing} patients are the ones with an initially misdiagnosed state of cancer \citep{cooperberg2011outcomes} or high-risk patients who choose AS instead of immediate treatment upon diagnosis. These are roughly 30\% of the population, having a cancer progression time less than 3.5 years. We label the remaining 20\% patients as \textit{intermediate progressing} patients. 

For \textit{fast progressing} patients (Panel~A,~Figure~\ref{Figure6}), we note that the personalized schedules with a fixed 10\% risk threshold and visit-specific threshold chosen using $\mbox{F}_1$ score, reduce one biopsy for 50\% of the patients, compared to PRIAS and annual schedule. Despite this, the delay (years) is similar for the personalized schedule with fixed 10\% risk threshold (median:~0.7,~IQR:~0.3--1.0), and the commonly used annual (median:~0.6,~IQR:~0.3--0.9) and PRIAS (median:~0.7,~IQR:~0.3--1.0) schedules.

For \textit{intermediate progressing} patients (Panel~A,~Figure~\ref{Figure6}), we note that the delay (years) due to personalized schedule with fixed 5\% risk threshold (median:~0.6,~IQR:~0.3--0.9) is comparable to that of annual schedule (median 0.5,~IQR:~0.2--0.7). However, it schedules fewer biopsies (median:~6,~IQR:~5--7) than the annual schedule (median:~7,~IQR:~5--8). The delay (years) for PRIAS (median:~0.7,~IQR:~0.3--1.3) and personalized schedule with fixed 10\% risk (median:~0.7,~IQR:~0.4--1.3) are similar, but the personalized approach schedules one less biopsy for 50\% of the patients. Although the approach with visit-specific risk threshold chosen using $\mbox{F}_1$ score schedules fewer biopsies than the 10\% fixed risk approach, it also has a higher delay.

The patients who are at the most advantage with the personalized schedules are the \textit{slow progressing} patients. These are a total of 50\% patients who did not progress during the entire study. Hence, the delay is not available for these patients (Panel~C of Figure~\ref{Figure6}). For all of these patients, annual schedule leads to 10 (unnecessary) biopsies. The schedule of the PRIAS program schedules a median of six biopsies (IQR:~4--8). In comparison, the biopsies scheduled by the personalized schedules using fixed 10\% risk threshold (median:~4,~IQR:~4--6) and visit-specific risk chosen using $\mbox{F}_1$ score (median:~2,~IQR:~2--4), are much fewer.

Overall, we observed that the personalized schedule which uses a 10\% risk threshold at all follow-up visits is dominant over the PRIAS schedule, biennial schedule of biopsies, and biopsies every one and a half years (see~Appendix~C for the latter two schedules). This personalized schedule not only schedules fewer biopsies than the aforementioned currently practiced schedules, but the delay in detection of cancer progression is also either equal or less. The personalized schedule which uses risk threshold chosen on the basis of classification accuracy ($\mbox{F}_1$ score) is dominant over the triennial schedule (see~Appendix~C) of biopsies. The personalized schedule which uses a 5\% risk threshold schedules fewer biopsies than the annual schedule, while the delay is only trivially more than the annual schedule.

\section{Discussion}
\label{sec:discussion}
We proposed a methodology which better balances the number of biopsies, and the delay in detection of cancer progression than the currently practiced biopsy schedules, for low-risk prostate cancer patients enrolled in active surveillance (AS) programs. The proposed methodology combines a patient's observed DRE and PSA measurements, and the time of the latest biopsy, into a personalized cancer progression risk function. If the cumulative risk of cancer progression at a follow-up visit is above a certain threshold, then a biopsy is scheduled. We conducted an extensive simulation study, based on a replica of the patients from the PRIAS program, to compare this personalized approach for biopsies with the currently practiced biopsy schedules. We found personalized schedules to be dominant over many of the current biopsy schedules (see \hyperref[sec:results]{Results}).

The main reason for the better performance of personalized schedules is that they account for the variation in cancer progression rate between patients, and also over time within the same patient. In contrast, the existing fixed/heuristic schedules ignore that roughly 50\% of the patients never progress in the first ten years of follow-up (\textit{slow progressing} patients) and do not require biopsies. The \textit{fast progressing} patients require early detection. However, existing methods of identifying these patients, such as the use of PSA doubling time in PRIAS, inappropriately assume that PSA evolves linearly over time. Thus, they may not correctly identify such patients. The personalized approach, however, models the PSA profiles non-linearly. Furthermore, it appends information from PSA with information from DRE and previous biopsy results and combines them into a single a cancer progression risk function. The risk function is a finer quantitative measure than individual data measurements observed for the patients. In comparison to decision making with flowcharts, the risk as a single measure of a patient's underlying state of cancer may facilitate shared decision making for biopsies.

Existing work on reducing the burden of biopsies in AS primarily advocates less frequent heuristic schedules of biopsies \citep{inoue2018comparative} (e.g., biopsies biennially instead of annually). To our knowledge, risk-based biopsy schedules have barely been explored yet in AS \citep{nieboer2018active,bruinsma2016active}. The part of our results pertaining to the fixed/heuristic schedules is comparable with corresponding results obtained in existing work \citep{inoue2018comparative}, even though the AS cohorts are not the same. Thus, we anticipate similar validity for the results pertaining to the personalized schedules.

A limitation of the personalized approach is that the choice of risk threshold is not straightforward, as different thresholds lead to different combinations of the number of biopsies and the delay in detection of cancer progression. An approach is to choose a risk threshold which leads to personalized schedule dominant (e.g., 10\% risk) over the currently practiced schedules, for a given delay. Since personalized biopsy schedules are less burdensome, they may lead to better compliance. A second limitation is that the results that we presented are valid only in a 10 year follow-up period, whereas prostate cancer is a slow progressing disease. Thus more detailed results, especially for \textit{slow progressing} patients cannot be estimated. However, very few AS cohorts have a longer follow-period than PRIAS \citep{bruinsma2016active}. In a screening setting often the ethno-racial background of the patient, as well as the history of cancer in first degree relatives are checked. Our model does not take into account either. The reason is that the history of cancer in relatives been found to be predictive of cancer progression only in African-American patients \citep{goh2013clinical,telang2017prostate}. This is also evident by the fact that PRIAS and many other surveillance programs do not utilize this information in their biopsy protocols \citep{bokhorst2016decade,nieboer2018active}. In addition, patients who have a higher risk of an aggressive form of cancer are usually not recommended active surveillance. Hence the proposed model is relevant only for low-risk prostate cancer patients eligible for active surveillance. An exception is the active surveillance patients who are old and/or have comorbid illnesses. Currently, such patients may be removed from active surveillance and are instead offered the less intensive watchful waiting \citep{bokhorst2016decade} option. It is also possible to model watchful waiting as a competing risk in our model. However, this falls outside the scope of the current work because cancer progression as detected via biopsy is the standard trigger for treatment advice. Lastly, our results are not valid when the patient data is missing not at random (MNAR).

There are multiple ways to extend the personalized decision making approach. For example, biopsy Gleason grading is susceptible to inter-observer variation \citep{coley2017}. Thus accounting for it in our model will be interesting to investigate further. To improve the decision making methodology, future consequences of a biopsy can be accounted for in the model by combining Markov decision processes with joint models for time-to-event and longitudinal data. There is also a potential for including diagnostic information from magnetic resonance imaging (MRI), such as the volume of the prostate tumor as a longitudinal measurement in our model. The resulting predictions can be used to the decide the time of next MRI as well as to make a decision of biopsy. The same holds true for the quality of life measures as well. However, given the scarceness of both MRI and quality of life measurements in the dataset, including them in the current model may not be feasible. We intend to further validate our results in a multi-center AS cohort, and subsequently develop a web application to assist in making shared decisions for biopsies.

\section*{Acknowledgments}
The first and last Authors would like to acknowledge support by Nederlandse Organisatie voor Wetenschappelijk Onderzoek (the national research council of the Netherlands) VIDI grant nr. 016.146.301, and Erasmus University Medical Center funding. The funding agreement ensured the authors’ independence in designing the study, interpreting the data, writing, and publishing the report. The Authors also thank the Erasmus University Medical Center's Cancer Computational Biology Center for giving access to their IT-infrastructure and software that was used for the computations and data analysis in this study. Lastly, we thank Joost Van Rosmalen from the Department of Biostatistics, Erasmus University Medical Center for feedback on the manuscript.

\section*{Supplementary material}
Supplementary material for this article, containing Appendix A--D are available at \url{https://github.com/anirudhtomer/prias/blob/master/report/decision_analytic/mdm/latex/supplementary.pdf}

\clearpage
\printbibliography

@article{pearson1994mixed,
  title={Mixed-effects regression models for studying the natural history of prostate disease},
  author={Pearson, Jay D and Morrell, Christopher H and Landis, Patricia K and Carter, H Ballentine and Brant, Larry J},
  journal={Statistics in Medicine},
  volume={13},
  number={5-7},
  pages={587--601},
  year={1994},
  publisher={Wiley Online Library}
}

@article{bruinsma2016active,
  title={Active surveillance for prostate cancer: a narrative review of clinical guidelines},
  author={Bruinsma, Sophie M and Bangma, Chris H and Carroll, Peter R and Leapman, Michael S and Rannikko, Antti and Petrides, Neophytos and Weerakoon, Mahesha and Bokhorst, Leonard P and Roobol, Monique J and Ehdaie, Behfar and others},
  journal={Nature Reviews Urology},
  volume={13},
  number={3},
  pages={151--167},
  year={2016},
  publisher={Nature Publishing Group}
}

@article{nieboer2018active,
  title={Active surveillance: a review of risk-based, dynamic monitoring},
  author={Nieboer, Daan and Tomer, Anirudh and Rizopoulos, Dimitris and Roobol, Monique J and Steyerberg, Ewout W},
  journal={Translational andrology and urology},
  volume={7},
  number={1},
  pages={106--115},
  year={2018},
  publisher={AME Publications}
}

@article{vickers2006decision,
  title={Decision curve analysis: a novel method for evaluating prediction models},
  author={Vickers, Andrew J and Elkin, Elena B},
  journal={Medical Decision Making},
  volume={26},
  number={6},
  pages={565--574},
  year={2006},
  publisher={Sage Publications Sage CA: Thousand Oaks, CA}
}

@article{fujita2009serial,
  title={Serial prostate biopsies are associated with an increased risk of erectile dysfunction in men with prostate cancer on active surveillance},
  author={Fujita, Kazutoshi and Landis, Patricia and McNeil, Brian K and Pavlovich, Christian P},
  journal={The Journal of urology},
  volume={182},
  number={6},
  pages={2664--2669},
  year={2009},
  publisher={Elsevier}
}

@article{coley2017,
title = {Prediction of the pathologic {Gleason} score to inform a personalized management program for prostate cancer},
journal = {European Urology},
volume = {72},
number = {1},
pages = {135--141},
year = {2017},
author = {Coley, Rebecca Yates and Zeger, Scott L and Mufaddal Mamawala and Pienta, Kenneth J and Carter, Herbert Ballentine}
}

@article{carvalho,
  title={Estimating the risks and benefits of active surveillance protocols for prostate cancer: a microsimulation study},
  author={de Carvalho, Tiago M and Heijnsdijk, Eveline AM and de Koning, Harry J},
  journal={BJU international},
  volume={119},
  number={4},
  pages={560--566},
  year={2017},
  publisher={Wiley Online Library}
}

@article{etzioni2002overdiagnosis,
  title={Overdiagnosis due to prostate-specific antigen screening: lessons from {US} prostate cancer incidence trends},
  author={Etzioni, Ruth and Penson, David F and Legler, Julie M and Di Tommaso, Dante and Boer, Rob and Gann, Peter H and Feuer, Eric J},
  journal={Journal of the National Cancer Institute},
  volume={94},
  number={13},
  pages={981--990},
  year={2002},
  publisher={Oxford University Press}
}

@article{inoue2018comparative,
  title={Comparative Analysis of Biopsy Upgrading in Four Prostate Cancer Active Surveillance Cohorts},
  author={Inoue, Lurdes YT and Lin, Daniel W and Newcomb, Lisa F and Leonardson, Amy S and Ankerst, Donna and Gulati, Roman and Carter, H Ballentine and Trock, Bruce J and Carroll, Peter R and Cooperberg, Matthew R and others},
  journal={Annals of internal medicine},
  volume={168},
  number={1},
  pages={1--9},
  year={2018},
  publisher={Am Coll Physicians}
}

@book{de1978practical,
  title={A practical guide to splines},
  author={De Boor, Carl},
  volume={27},
  year={1978},
  publisher={Springer-Verlag New York}
}

@article{loeb2014heterogeneity,
  title={Heterogeneity in active surveillance protocols worldwide},
  author={Loeb, Stacy and Carter, H Ballentine and Schwartz, Mark and Fagerlin, Angela and Braithwaite, R Scott and Lepor, Herbert},
  journal={Reviews in urology},
  volume={16},
  number={4},
  pages={202--203},
  year={2014},
  publisher={MedReviews, LLC}
}

@article{eilers1996flexible,
title={Flexible smoothing with {B-splines} and penalties},
author={Eilers, Paul HC and Marx, Brian D},
journal={Statistical Science},
pages={89--121},
year={1996},
volume={11},
number={2},
publisher={Institute of Mathematical Statistics}
}

@article{cooperberg2011outcomes,
  title={Outcomes of active surveillance for men with intermediate-risk prostate cancer},
  author={Cooperberg, Matthew R and Cowan, Janet E and Hilton, Joan F and Reese, Adam C and Zaid, Harras B and Porten, Sima P and Shinohara, Katsuto and Meng, Maxwell V and Greene, Kirsten L and Carroll, Peter R},
  journal={Journal of Clinical Oncology},
  volume={29},
  number={2},
  pages={228--234},
  year={2011},
  publisher={American Society of Clinical Oncology}
}

@article{lin2000latent,
  title={A latent class mixed model for analysing biomarker trajectories with irregularly scheduled observations},
  author={Lin, Haiqun and McCulloch, Charles E and Turnbull, Bruce W and Slate, Elizabeth H and Clark, Larry C},
  journal={Statistics in Medicine},
  volume={19},
  number={10},
  pages={1303--1318},
  year={2000},
  publisher={Wiley Online Library}
}

@article{schroder1992tnm,
  title={The {TNM} classification of prostate cancer},
  author={Schr{\"o}der, FH and Hermanek, P and Denis, L and Fair, WR and Gospodarowicz, MK and Pavone-Macaluso, M},
  journal={The Prostate},
  volume={21},
  number={S4},
  pages={129--138},
  year={1992},
  publisher={Wiley Online Library}
}

@article{GlobalCancerStats2012,
title={Global cancer statistics, 2012},
author={Torre, Lindsey A and Bray, Freddie and Siegel, Rebecca L and Ferlay, Jacques and Lortet-Tieulent, Joannie and Jemal, Ahmedin},
journal={CA: A Cancer Journal for Clinicians},
volume={65},
number={2},
pages={87--108},
year={2015},
publisher={Wiley}
}

@article{ehdaie2014impact,
  title={The impact of repeat biopsies on infectious complications in men with prostate cancer on active surveillance},
  author={Ehdaie, Behfar and Vertosick, Emily and Spaliviero, Massimiliano and Giallo-Uvino, Anna and Taur, Ying and O'sullivan, Maryellen and Livingston, Jennifer and Sogani, Pramod and Eastham, James and Scardino, Peter and others},
  journal={The Journal of urology},
  volume={191},
  number={3},
  pages={660--664},
  year={2014},
  publisher={Elsevier}
}

@article{bokhorst2016decade,
title={A decade of active surveillance in the {PRIAS} study: an update and evaluation of the criteria used to recommend a switch to active treatment},
author={Bokhorst, Leonard P and Valdagni, Riccardo and Rannikko, Antti and Kakehi, Yoshiyuki and Pickles, Tom and Bangma, Chris H and Roobol, Monique J and {\relax PRIAS study group}},
journal={European Urology},
volume={70},
number={6},
pages={954--960},
year={2016},
publisher={Elsevier}
}

@article{goh2013clinical,
  title={Clinical implications of family history of prostate cancer and genetic risk single nucleotide polymorphism (SNP) profiles in an active surveillance cohort},
  author={Goh, Chee L and Saunders, Edward J and Leongamornlert, Daniel A and Tymrakiewicz, Malgorzata and Thomas, Karen and Selvadurai, Elizabeth D and Woode-Amissah, Ruth and Dadaev, Tokhir and Mahmud, Nadiya and Castro, Elena and others},
  journal={BJU international},
  volume={112},
  number={5},
  pages={666--673},
  year={2013},
  publisher={Wiley Online Library}
}

@article{telang2017prostate,
  title={Prostate cancer family history and eligibility for active surveillance: a systematic review of the literature},
  author={Telang, Jaya M and Lane, Brian R and Cher, Michael L and Miller, David C and Dupree, James M},
  journal={BJU international},
  volume={120},
  number={4},
  pages={464--467},
  year={2017},
  publisher={Wiley Online Library}
}

@article{bokhorst2015compliance,
title={Compliance rates with the {Prostate Cancer Research International Active Surveillance (PRIAS)} protocol and disease reclassification in noncompliers},
author={Bokhorst, Leonard P and Alberts, Arnout R and Rannikko, Antti and Valdagni, Riccardo and Pickles, Tom and Kakehi, Yoshiyuki and Bangma, Chris H and Roobol, Monique J and {\relax PRIAS study group}},
journal={European Urology},
volume={68},
number={5},
pages={814--821},
year={2015},
publisher={Elsevier}
}

@article{tsiatis2004joint,
title={Joint modeling of longitudinal and time-to-event data: an overview},
author={Tsiatis, Anastasios A and Davidian, Marie},
journal={Statistica Sinica},
volume={14},
number={3},
pages={809--834},
year={2004},
publisher={Institute of Statistical Science, Academia Sinica}
}

@article{rizopoulosJMbayes,
author = {Dimitris Rizopoulos},
title = {The {R} Package {JMbayes} for Fitting Joint Models for Longitudinal and Time-to-Event Data Using {MCMC}},
journal = {Journal of Statistical Software},
volume = {72},
number = {7},
year = {2016},
issn = {1548-7660},
pages = {1--46},
publisher={Foundation for Open Access Statistics}
}

@article{landmarking2017,
  title={Dynamic predictions with time-dependent covariates in survival analysis using joint modeling and landmarking},
  author={Rizopoulos, Dimitris and Molenberghs, Geert and Lesaffre, Emmanuel MEH},
  journal={Biometrical Journal},
  volume={59},
  number={6},
  pages={1261--1276},
  year={2017},
  publisher={Wiley Online Library}
}

@article{rizopoulos2011dynamic,
title={Dynamic Predictions and Prospective Accuracy in Joint Models for Longitudinal and Time-to-Event Data},
author={Rizopoulos, Dimitris},
journal={Biometrics},
volume={67},
number={3},
pages={819--829},
year={2011},
publisher={Wiley}
}

@book{rizopoulos2012joint,
title={Joint Models for Longitudinal and Time-to-Event Data: With Applications in R},
author={Rizopoulos, Dimitris},
year={2012},
publisher={CRC Press},
isbn = {9781439872864}
}

\end{document}